# Akceleracja obliczeń algebry liniowej z wykorzystaniem masywnie równoległych, wielordzeniowych procesorów GPU
Świerczewski Ł.

## *Wprowadzenie do koncepcji budowy akceleratorów graficznych*

Pierwsze procesory graficzne umożliwiające wykonywanie obliczeń związanych nie tylko z generowanym obrazem pojawiły się w kwietniu 2006 roku. Od tego momentu wydajne akceleratory są niezwykle dynamicznie rozwijającą się gałęzią rynku, na którą z wielkim zainteresowaniem spoglądają nie tylko cywilni, indywidualni użytkownicy ale także jednostki badawcze z całego świata.

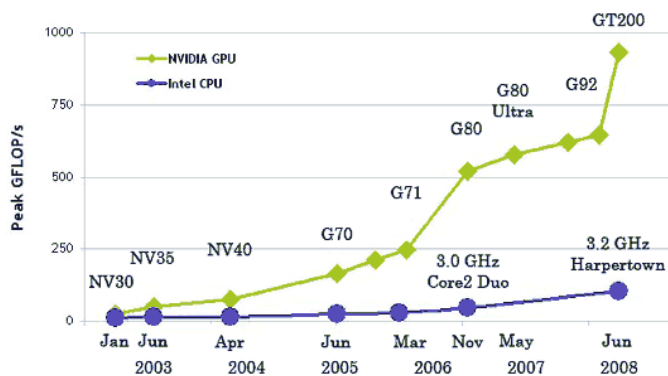

**Rysunek 1. Dynamiczny rozwój możliwości obliczeniowych kolejnych generacji GPU w porównaniu do CPU.**
Źródło: Materiały techniczne firmy nVidia

Główną różnicą pomiędzy GPU, a CPU jest różna ich klasyfikacja w taksonomii Flynna. Współczesne procesory wielordzeniowe opierają się na architekturze MIMD (ang. Multiple Instruction, Multiple Data). Mogą więc równolegle wykonywać wiele różnych zadań, z których każde posiada własny strumień danych. Zrównoleglenie w tym przypadku występuje zarówno na poziomie instrukcji, jak i danych. Akceleratory graficzne są jednak w znacznie większym stopniu ograniczone do architektury SIMD (ang. Single Instruction, Multiple Data), według założeń której pojedynczy strumień rozkazów przetwarza wiele strumieni danych. GPU Jest zbudowany z wielu procesorów strumieniowych, mających dostęp do rejestrów i pamięci podręcznej w znaczącym stopniu ograniczona w stosunku do CPU. Struktura jednostek obliczeniowych jest jednak w tym przypadku znacznie bardziej zaawansowana - najnowsze modele kart graficznych mogą posiadać nawet 512 rdzeni co czyni z nich układy masywnie równoległe. Różnicę tą przedstawiono na Rys. 2. Model ten wprowadza pewne restrykcje - obliczenia realizowane w ramach jednego strumienia nie mogą w bezpośredni sposób wpłynąć na operacje prowadzone w innych potokach lub nawet w tym samym potoku ale dla innych danych wejściowych.

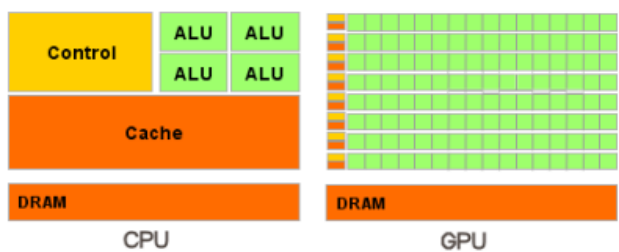

**Rysunek 2. Porównanie architektury GPU z CPU.**
Źródło: Materiały techniczne firmy nVidia

Układy graficzne posiadają własną, wyspecjalizowaną hierarchię pamięci. W przypadku CPU mamy dostęp do maksymalnie trzech poziomów pamięci podręcznej (ang. cache memory) różniącej się praktycznie jedynie poziomem przeźroczystości, rozmiarem i czasem dostępu. Procesor klasyczny nie posiada dostępu do wszystkich sekcji pamięci znajdujących się na karcie graficznej. CPU najczęściej jedynie wykonuje operację skopiowania wymaganych informacji z pamięci RAM do pamięci globalnej karty graficznej (największy ale i najwolniejszy segment pamięci) i wykonuje wywołanie kernela - funkcji wykonywanej przez akcelerator. Pozostałe zadania realizowane są jedynie z wykorzystaniem GPU, który angażuje zasoby sprzętowe (rejestry i pamięć niskopoziomową), do których tylko on posiada dostęp. Po przeprowadzeniu wymaganych obliczeń przez procesory strumieniowe dane wyjściowe zostają umieszczone w pamięci globalnej, z której mogą być pobrane przez CPU. Słabym punktem obliczeń wykonywanych na karcie graficznej jest przesył danych na linii: pamięć RAM - akcelerator graficzny. Jak się jednak okazuje komunikacja często wcale znacząco nie zmniejsza efektywności obliczeń.

Procesory strumieniowe są pogrupowane po 8 lub 32 (w zależności od generacji) w większe bloki organizacyjne nazwane muliprocesorami. Multiprocesory posiadają znacznie większą autonomię - niezależną pamięć przeznaczoną na dane współdzielone (ang. shared memory), stałe (ang. constant cache) i tekstury (ang. texture cache). Karty starszej generacji Tesla C1060 posiadają 30 multiprocesorów po 8 procesorów strumieniowych w każdym. Organizację pamięci oraz jednostek obliczeniowych przedstawia Rys. 3. oraz Tab. 1.

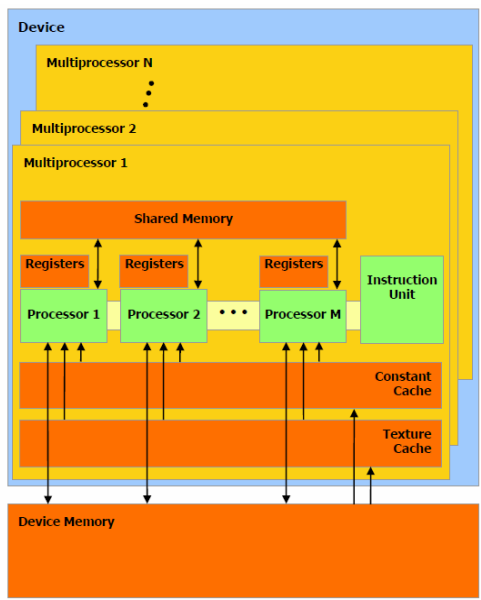

**Rysunek 3. Schemat budowy rdzenia opartego o architekturę CUDA z uwzględnieniem różnego poziomu pamięci.**
Źródło: Materiały techniczne firmy nVidia

| Typ pamięci | Zintegrowana z GPU | Czas dostępu | Rozmiar | Rodzaje dostępu |
|---|---|---|---|---|
| Register | Tak | Niski | Mały | GPU(Read, Write) CPU(Brak) |
| Shared | Tak | Niski | Mały | GPU(Read, Write) CPU(Brak) |
| Constant | Nie | Średni | Średni | GPU(Read) CPU(Write) |
| Global (Device Memory) | Nie | Duży | Duży | GPU(Read, Write) CPU(Read, Write) |

**Tabela 1. Hierarchia pamięci dostępna do wykorzystania przez programistę.**
Źródło: Opracowanie własne

*Implementacje algorytmów*

Standardowy sposób mnożenia macierzy nazywamy mnożeniem Cauchy'ego. Jeżeli C = A · B definiuje iloczyn macierzy A i B, a $c_{i,j}$ oznacza element w macierzy C znajdujący się na pozycji (i, j) to:

$$c_{i,j}=\sum_{r=1}^{m} a_{i,r}b_{r,j}=a_{i,1}b_{1,j}+a_{i,2}b_{2,j}+...+a_{i,m}b_{m,j}$$

dla każdej pary i,j dla której $1 \leq i \leq n$ oraz $1 \leq j \leq p$ .

Naiwny algorytm mnożenia macierzy o rozmiarze x·y przez macierz y·z ma złożoność O(xyz). W przypadku macierzy kwadratowych uzyskujemy złożoność $O(n^3)$. Pierwszy z wydajniejszych algorytmów obliczających iloczyn macierzy opracował Volker Strassen. Jego złożoność wynosi $O(n^{2,807})$. Nie znajduje on jednak często zastosowania ze względu na brak numerycznej stabilności. Istnieją także wyspecjalizowane rozwiązania dla macierzy posiadających szczególne cechy. Prosty algorytm mnożenia macierzy diagonalnych mieści się w klasie złożoności O(n). Przykładową realizację mnożenia macierzy w języku C przedstawiono w Listing 1.

```
for(i = 0; i < rows_size_A; i++) {
    for(j = 0; j < columns_size_B; j++) {
        matrix_C[i_columns_size_B] = 0;
        k_ columns_size_B = j;
        for(k = 0; k < columns_size_A; k++) {
            matrix_C[i_columns_size_B] +=
            matrix_A[k + i_columns_size_A] * matrix_B[k_columns_size_B];
            k_ columns_size_B += colums_size_B;
        }
        i_columns_size_B++;
    }
    i_columns_size_A += columns_size_A;
}
```

**Listing 1. Sekwencyjny algorytm mnożenia macierzy w języku C.**
Źródło: Opracowanie własne

Najprostsza wersja algorytmu wykonywana na akceleratorze graficznym może wyglądać tak jak na Listingu 2. Jak widać kod bazuje na standardzie ANSI C z dodatkowymi rozszerzeniami CUDA. Specyficzne słowo kluczowe __global__ definiuje funkcję będącą jądrem. W momencie wywołania jądra przez CPU zostaje wygenerowana siatka wątków na karcie graficznej. Wszystkie wątki wykonują ten sam kod, a więc musi istnieć mechanizm umożliwiający wyodrębnienie określonego wątku i wskazanie na jakim zakresie danych ma on pracować. Umożliwiają to stałe threadIdx.x oraz threadIdx.y. Definiują one lokalizację wątku na dwuwymiarowej siatce za pomocą odciętej x i rzędnej y w typowym układzie współrzędnych kartezjańskich. Wątki są umieszczone w blokach, a wszystkie bloki składają się na grid. Dla uproszczenia możemy przyjąć, że wywołujemy grid złożony z jednego bloku o rozmiarze 3. Mnożeniu będziemy poddawać macierz o rozmiarze 3x3. Zostanie więc wygenerowana siatka wątków 3x3, gdzie każdy wątek będzie miał określone współrzędne x i y. Dzięki temu możliwe będzie równoległe obliczenie każdego elementu w macierzy wynikowej. Wątek thread_x = 0 i thread_y = 1 będzie więc odpowiedzialny za wyliczenie elementu $C_{0,1}$. Bardzo małą ilość wątków w tym przykładzie ustalono tylko ze względu na prostą ilustrację zagadnienia. W praktyce grid składa się z wielu bloków, a pojedynczy blok może mieć rozmiar równy nawet 1024. Strukturę logiczną kerneli wykonywanych przez GPU przedstawiono na Rys. 4.

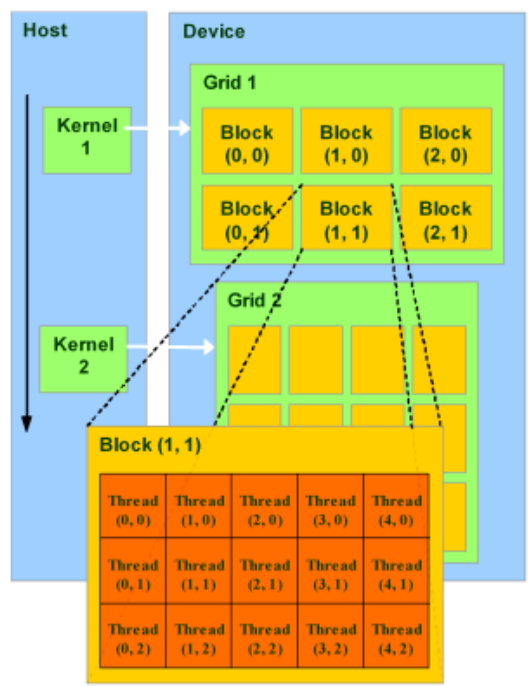

**Rysunek 4. Struktura logiczna kerneli w architekturze CUDA.**
Źródło: Materiały techniczne firmy nVidia

```
__global__ void matrixMul( float *matrix_C, float *matrix_A, float *matrix_B, int size_A, int size_B)
{
        unsigned int thread_x = threadIdx.x;
        unsigned int thread_y = threadIdx.y;

        float value_temporary = 0;
        for (int i = 0; i < size_A; ++i)
        {
                value_temporary += matrix_A[thread_y * size_A + i] * matrix_B[i * size_B + thread_x];
        }

        matrix_C[thread_y * size_A + thread_x] = value_temporary;
}
```

**Listing 2. Najprostsza wersja mnożenia macierzy na GPU.**
Źródło: Opracowanie własne

Wyżej przedstawiony algorytm nie umożliwi nam mnożenia macierzy większych niż 1024x1024, ponieważ rozmiar bloku na współczesnych akceleratorach nie może przekraczać 1024. Wszelkie dalsze przykłady będę opierał ma macierzach 4096x4096. W tym przypadku wątki będziemy musieli już podzielić na większą ilość bloków. Zakładając np., że chcemy aby blok miał rozmiar równy 16 (więc mieścił 256 wątków) wiemy, że musimy posiadać $4096 \div 256 = 16$ bloków.

Stara metoda określania zakresu na jakich operował dany wątek po wprowadzeniu nowych zmian jest nieprawidłowa. Współrzędne thread_x i thread_y powinny teraz być wyznaczane z uwzględnieniem numeru oraz rozmiaru bloku. Poprawiony algorytm przedstawia Listing 3.

```
__global__ void matrix_mul_cuda( float *matrix_C, float *matrix_A, float *matrix_B, unsigned
long int size_A, unsigned long int size_B)
{
    unsigned long int thread_x = blockIdx.x * BLOCK_SIZE + threadIdx.x;
    unsigned long int thread_y = blockIdx.y * BLOCK_SIZE + threadIdx.y;

    float value_temporary = 0;
    for (unsigned int i = 0; i < size_A; ++i)
    {
        value_temporary += matrix_A[thread_y * size_A + i] * matrix_B[i * size_B + thread_x];
    }

    matrix_C[thread_y * size_A + thread_x] = value_temporary;
}
```

**Listing 3. Nieoptymalna wersja mnożenia macierzy na GPU.**
Źródło: Opracowanie własne

Przedstawiony sposób nie jest jednak nadal optymalny. W Tab. 1 przedstawiono hierarchię pamięci GPU. We wszystkich wcześniejszych implementacjach dla akceleratorów graficznych korzystaliśmy zawsze z pamięci globalnej, która jest najwolniejsza. Powinniśmy więc jak najbardziej ograniczyć wykorzystanie tego segmentu danych względem dużo szybszej pamięci współdzielonej. Pamięć współdzielona ma jednak znacznie ograniczony rozmiar i nie możemy w niej umieścić wszystkich danych. Jest ona wspólna w ramach jednego multiprocesora. Multiprocesor w danym momencie wykonuje jeden blok wątków. W jednym bloku będziemy potrzebować dwóch submacierzy o rozmiarze 16x16 (gdy rozmiar bloku = 16), w których zostaną przechowane wybrane elementy z macierzy A i B. W naszym przypadku rozmiar pamięci współdzielonej przydzielonej dla jednego wątku będzie więc wynosił $2 \cdot 16 \cdot 16 \cdot 8\ B = 4\ KB$, gdy pod uwagę brane są zmienne typu double lub complex_float. Jest to wielkość akceptowalna, ponieważ nie przekracza wartości maksymalnej wynoszącej 16 (starsze akceleratory) lub 48 KB (nowsze akceleratory).

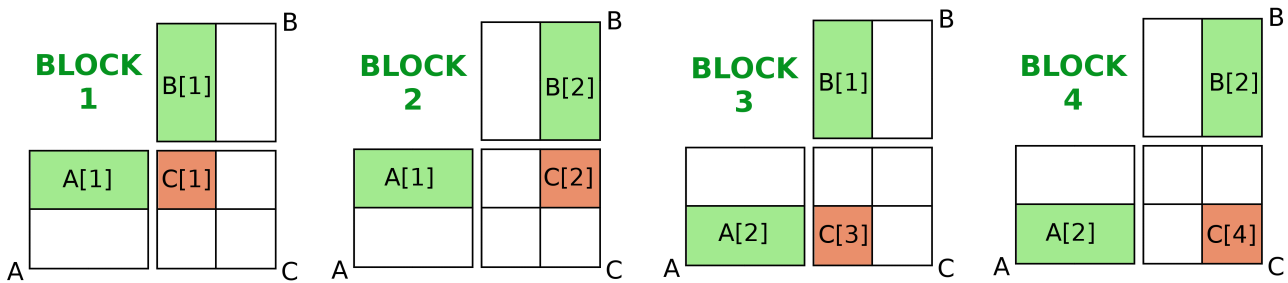

**Rysunek 5. Podział danych między bloki.**
Źródło: Opracowanie własne

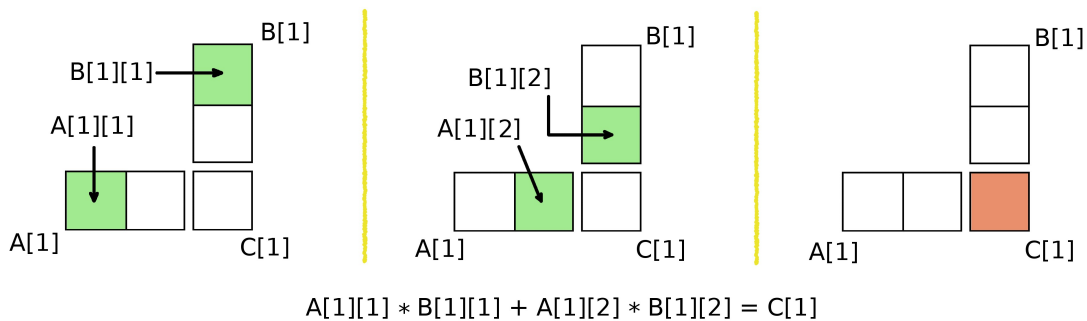

$A[1][1] * B[1][1] + A[1][2] * B[1][2] = C[1]$

**Rysunek 6. Uproszczona realizacja operacji w ramach jednego bloku.**
Źródło: Opracowanie własne

Przykładowy podział wykonywanych zadań przez bloki został zaprezentowany na Rys. 5. Na zielono zaznaczono elementy macierzy A i B wymagane do obliczenia czerwonego fragmentu w macierzy C. Początkowo zostają one przeniesione do pamięci współdzielonej dzięki czemu podczas właściwego mnożenia elementów możliwy jest znacznie szybszy dostęp. W przypadku obliczania iloczynu submacierzy w ramach bloku postępujemy w sposób intuicyjny. Przedstawia to Rys. 6. Macierz wynikowa jest ostatecznie kopiowana do pamięci globalnej skąd może zostać przez procesor CPU skopiowana do pamięci operacyjnej RAM. Implementację całego algorytmu zaprezentowano w Listing 4.

```
__global__ void matrixMul( float *matrix_C, float *matrix_A, float *matrix_B, unsigned long int size_A, unsigned long int size_B)
{
        unsigned int block_x = blockIdx.x;
        unsigned int block_y = blockIdx.y;

        unsigned int thread_x = threadIdx.x;
        unsigned int thread_y = threadIdx.y;

        unsigned int A_start = size_A * BLOCK_SIZE * block_y;

        unsigned int A_stop  = A_start + size_A - 1;

        unsigned int A_step = BLOCK_SIZE;

        unsigned int B_start = BLOCK_SIZE * block_x;

        unsigned int B_step  = BLOCK_SIZE * size_B;

        float C_temporary;
        C_temporary = 0;

        for (unsigned int a = A_start, b = B_start; a <= A_stop; a += A_step, b += B_step)
        {
                __shared__ float A_shared[BLOCK_SIZE][BLOCK_SIZE];
                __shared__ float B_shared[BLOCK_SIZE][BLOCK_SIZE];

                A_shared[thread_y][thread_x] = matrix_A[a + size_A * thread_y + thread_x];
                B_shared[thread_y][thread_x] = matrix_B[b + size_B * thread_y + thread_x];

                __syncthreads();

                for (unsigned int k = 0; k < BLOCK_SIZE; ++k)
                {
                        C_temporary += A_shared[thread_y][k] * B_shared[k][thread_x];
                }
                __syncthreads();
        }

        int c = size_B * BLOCK_SIZE * block_y + BLOCK_SIZE * block_x;
        matrix_C[c + size_B * thread_y + thread_x] = C_temporary;
}
```

**Listing 4. Zoptymalizowana wersja mnożenia macierzy na GPU wykorzystująca shared memory.**
Źródło: Opracowanie własne

Zastosowanie powyższego schematu postępowania może także umożliwić podział zadań pomiędzy różne procesory graficzne. Dla przykładu moduł nVidia Tesla S2050 złożony z czterech kart Tesla C2050 umożliwi teoretycznie czterokrotne zwiększenie wydajności względem jednej karty. Jednak w przypadku mnożenia macierzy musiałyby być one bardzo duże aby zastosowanie kilku akceleratorów przyniosło wymierne korzyści.

### *Rezultaty i wnioski*

Podczas testów wydajnościowych zauważono drastyczny wzrost wydajności w przypadku wykorzystania procesorów GPU. Najwyższe przyśpieszenie zaobserwowano na karcie Tesla C2050 w przypadku typu float. Wynosiło ono 1195,13 co oznacza, że rezultaty sekundy pracy akceleratora są równoważne wynikom zwróconym po ok. 20 minutach przez algorytm sekwencyjny wykonywany na klasycznym procesorze CPU. Porównanie wydajności różnych GPU z CPU przedstawiono na Rys. 7.

Analizie poddano także algorytm dodawania / odejmowania macierzy. Ze względu na niską złożoność obliczeniową problemu nie zauważono jednak w tym przypadku większych zysków. Dodawanie macierzy typów prostych float lub double o rozmiarze 4096x4096 wymaga wykonania tylko 16 777 216 operacji elementarnych. Procesor CPU wykonuje to zadanie w około 0,1 sekundy z całkowitym narzutem około 200 000 000 operacji (dla zmiennych float). Dokładna liczba operacji została wyznaczona na podstawie wskazań sprzętowego licznika rozkazów wbudowanego w procesory marki Intel, a pomiaru dokonano za pomocą wstawki asemblerowej wywołanej z poziomu języka C. W przypadku jedynie wywołania jądra na karcie graficznej ilość wykonanych operacji przez CPU spada do granicy 200 000. Obydwa przypadki charakteryzują czasy wykonywania tak niskie, że różnica jest praktycznie niemierzalna. Dokładniejsze wyniki przedstawia Rys. 9.

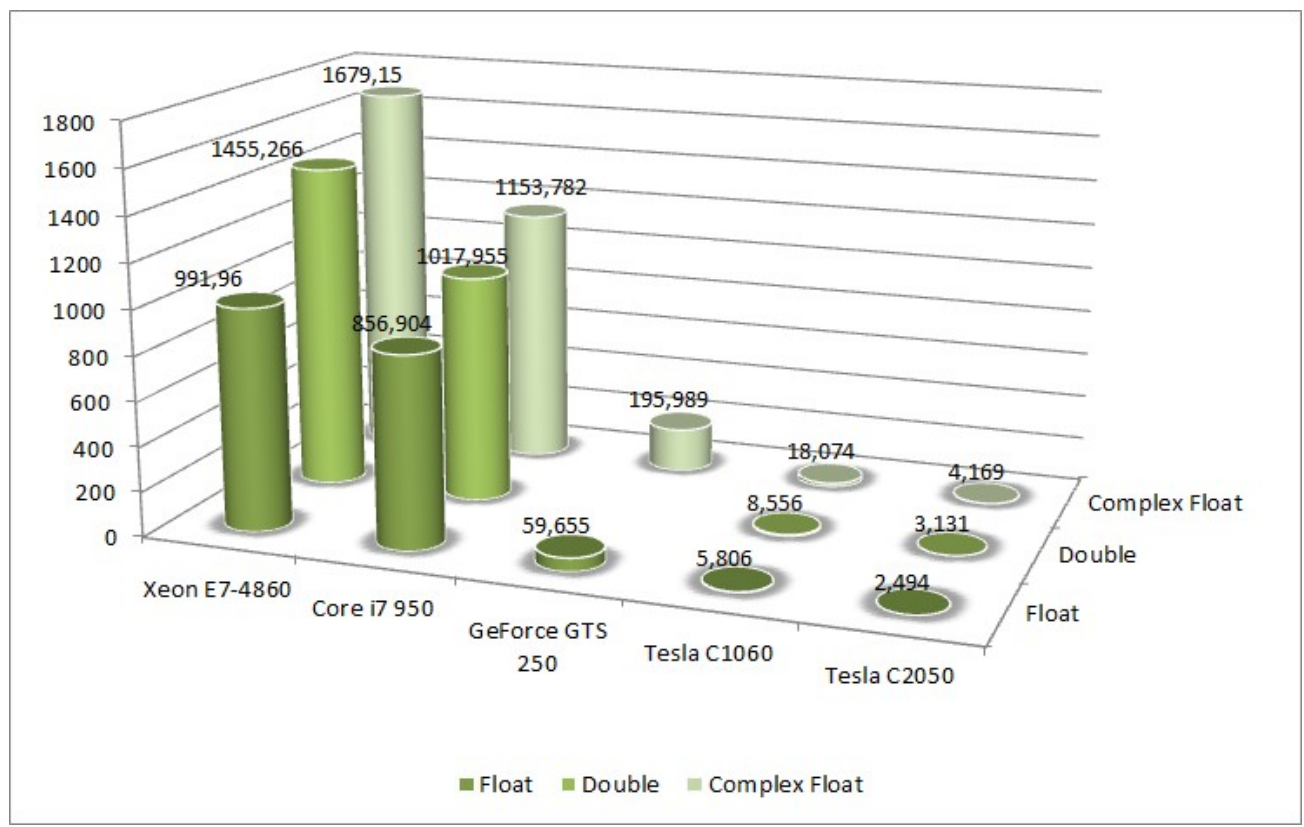

**Rysunek 7. Czas realizacji (w sekundach) algorytmu mnożenia macierzy z wykorzystaniem różnych układów.**
Źródło: Opracowanie własne

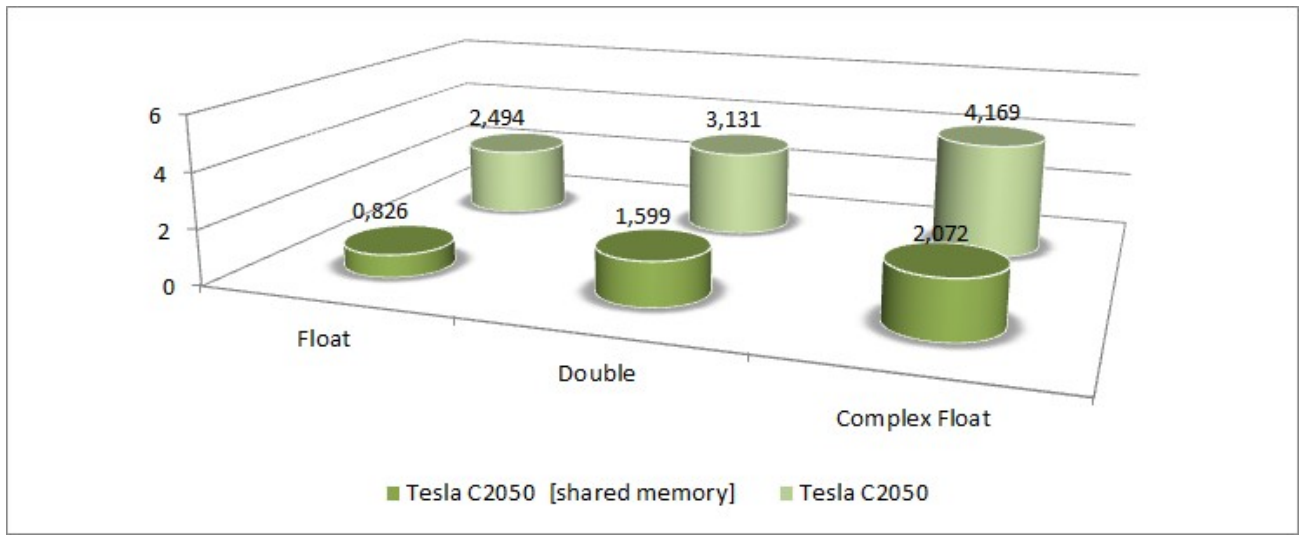

**Rysunek 8. Porównanie czasów realizacji algorytmu mnożenia macierzy ze wsparciem i bez wsparcia dla shared memory.**
Źródło: Opracowanie własne

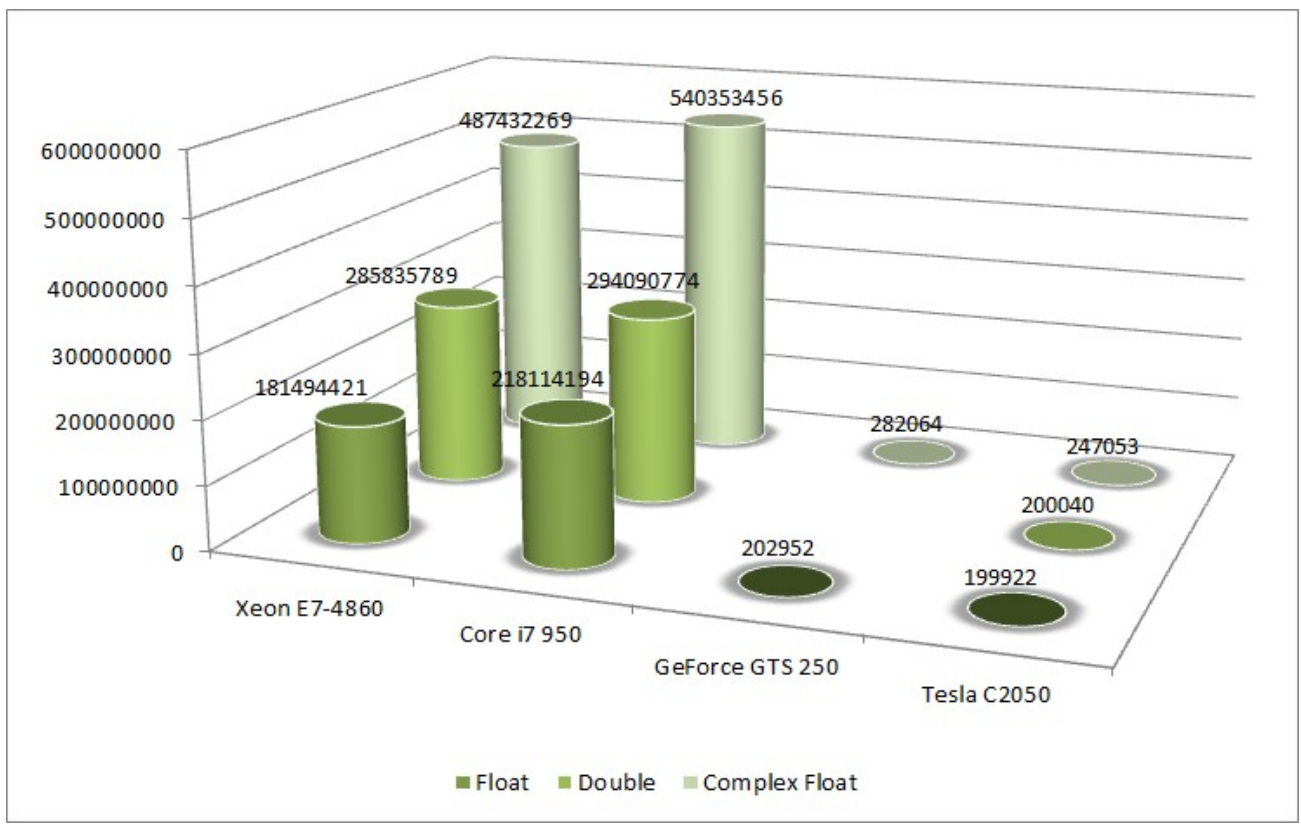

**Rysunek 9. Czas realizacji (ilość operacji procesora CPU) algorytmu dodawania / odejmowania macierzy z wykorzystaniem różnych układów.**
Źródło: Opracowanie własne

|  | Float | Przyśpieszenie* | Double | Przyśpieszenie* | Complex Float | Przyśpieszenie* |
|---|---|---|---|---|---|---|
| Intel Xeon E7-4860 | 991,96 | 1 | 1455,27 | 1 | 1679,15 | 1 |
| Intel Core i7 950 | 856,9 | 1,16 | 1017,96 | 1,43 | 1153,78 | 1,46 |
| nVidia GeForce GTS 250 | 59,66 | 16,63 | - | - | 195,99 | 8,57 |
| nVidia Tesla C1060 | 5,81 | 170,85 | 8,56 | 170,09 | 18,07 | 92,9 |
| nVidia Tesla C2050 | 2,49 | 397,74 | 3,13 | 464,79 | 4,17 | 402,77 |
| nVidia Tesla C2050 [shared memory] | 0,83 | 1195,13 | 1,6 | 909,54 | 2,07 | 811,18 |

*Względem procesora Intel Xeon E7-4860

**Tabela 2. Prezentacja czasów wykonywania algorytmu mnożenia macierzy (w sekundach) oraz uzyskanych przyśpieszeń.**
Źródło: Opracowanie własne

Zaprezentowane wyniki świadczą o dużej sprawności układów graficznych, które są z pewnością w niektórych przypadkach świetną alternatywą dla konwencjonalnych metod obliczeniowych. Jest to szczególnie zauważalne w przypadku algorytmów mnożenia macierzy, które można łatwo zrównoleglić. Ciekawe efekty może przynieść także implementacja różnych schematów rozwiązywania układów równań nieliniowych - np. metody eliminacji Gaussa.

**Streszczenie**

Praca przedstawia aspekt wykorzystania współczesnych akceleratorów graficznych wspierających technologię CUDA na potrzeby wydajnych obliczeń z dziedziny algebry liniowej. W pełni programowalne karty graficzne są dostępne już od kilku lat zarówno dla zwykłych użytkowników, jak i jednostek naukowych. Umożliwiają one wykonywanie praktycznie dowolnych obliczeń charakteryzujących się wysoką wydajnością, która jest często nie do osiągnięcia na klasycznych procesorach CPU. Architektura GPU także w przypadku klasycznych problemów dotyczących algebry liniowej, która stanowi podstawę wielu obliczeń potrafi przynieść programiście szereg korzyści. Zaobserwowane podczas mnożenia macierzy przyśpieszenie na układzie nVidia Tesla C2050 wynoszące ponad 1000x względem zwykłego procesora jest wynikiem drastycznie redukującym czas oczekiwania na niektóre wyniki, a zarazem koszty ich uzyskania.

**Abstract**

The paper presents the aspect of use of modern graphics accelerators supporting CUDA technology for high-performance computing in the field of linear algebra. Fully programmable graphic cards have been available for several years for both ordinary users and research units. They provide the capability of performing virtually any computing with high performance, which is often beyond the reach of conventional CPUs. GPU architecture, also in case of classical problems of linear algebra which is the basis for many calculations, can bring many benefits to the developer. Performance increase, observed during matrix multiplication on nVidia Tesla C2050, was more than thousandfold compared to ordinary CPU, resulting in drastic reduction of latency for some of the results, thus the cost of obtaining them.

**Nota o autorze**

Łukasz Świerczewski, Instytut Informatyki i Automatyki, Państwowa Wyższa Szkoła Informatyki i Przedsiębiorczości w Łomży. Zainteresowania to głównie programowanie równoległe i matematyka teoretyczna. Strona internetowa: www.goldbach.pl/~lswierczewski email: luk.swierczewski@gmail.com